\documentclass[twocolumn,prc,preprintnumbers,superscriptaddress,floatfix,article]{revtex4}
\usepackage{dcolumn}
\usepackage{bm}
\usepackage{longtable}
\usepackage{graphicx,epsfig,latexsym,amssymb}
\usepackage{multirow,amsmath,array,booktabs,color}
\usepackage[section]{placeins}
\usepackage{CJK}

\begin{document}

\title{Predication of novel effects in rotational nuclei at high speed}
\author{Jian-You Guo}
\email[Corresponding author:]{jianyou@ahu.edu.cn}
\affiliation{School of physics and Optoelectronic Engineering, Anhui University, Hefei
230601, China}
\date{\today }

\begin{abstract}
The study of high-speed rotating matter is a crucial research topic in
physics due to the emergence of novel phenomena. In this paper, we combined
cranking covariant density functional theory (CDFT) with a similar
renormalization group approach to decompose the Hamiltonian from the
cranking CDFT into different Hermit components, including the
non-relativistic term, the dynamical term, the spin-orbit coupling, and the
Darwin term. Especially, we obtained the rotational term, the term relating
to Zeeman effect-like, and the spin-rotation coupling due to consideration of
rotation and spatial component of vector potential. By exploring these
operators, we aim to identify novel phenomena that may occur in rotating
nuclei. Signature splitting, Zeeman effect-like, spin-rotation coupling, and spin
current are among the potential novelties that may arise in rotating nuclei.
Additionally, we investigated the observability of these phenomena and their
dependence on various factors such as nuclear deformation, rotational
angular velocity, and strength of magnetic field.
\end{abstract}

\maketitle

%\begin{CJK*}{GBK}{song}

\section{Introduction}

The investigation of the structure and property of matter under extreme
conditions is an important research subject for physicists. With the
development of accelerator facilities, particle detector systems, and
techniques of nuclear spectroscopy, physicists have discovered many novel
phenomena with unexpected features in atomic nuclei. These novel phenomena
include neutron (proton) halos~\cite{Tanihata2013,Nakamura2017}, energy
level inversion~\cite{Wilson2016,Longfellow2020}, change of magic numbers~%
\cite{Ozawa2000,Ruiz2016,Leis2021}, exotic radioactivity~\cite%
{Olsen2013,Ma2015}, high spin states~\cite{Twin1986,Firestone2002}, and
others

The high-spin states provide an important opportunity to investigate the
microscopic mechanism of deformed shell structure, single-particle motion,
and collective excitation. Since the discovery of backbending phenomenon in
high spin states~\cite{Johnson1971}, the rotation excitation has become an
active frontier. The backbending~\cite{Johnson1971,Lee1977}, band
termination \cite{Bengtsson1983,Afanasjev1999}, signature inversion~\cite%
{Bengtsson1984}, superdeformed rotation~\cite{Twin1986,Firestone2002},
magnetic and antimagnetic rotations~\cite{Frauendorf1994, Frauendorf2001},
chiral rotation~\cite{Frauendorf1997} have attracted worldwide attention.

To understand these novel phenomena associated with rotational excitation,
physicists have developed various theoretical models, such as the cranked
Nilsson-Strutinsky method~\cite{Andersson1976}, the cranked shell model~\cite%
{Bengtsson1979}, the projected shell model~\cite{Hara1995}, the tilted-axis
cranking model~\cite{Frauendorf2001}, large-scale shell model calculations~%
\cite{Caurier2007}, and density functional theories~\cite%
{Afanasjev2000,Sakai2020}. However, these models have limitations due to the
breaking of time-reversal symmetry by the Coriolis operator and unpaired
nucleons, which results in time-odd components that are difficult to treat
in non-relativistic framework.

The self-consistent cranking covariant density functional theory (CDFT) has
successfully described the rotational excitations of atomic nuclei~\cite%
{Ideguchi2001,Ray2016}, including the magnetic rotation, antimagnetic
rotation, chiral doublet bands~\cite{Frauendorf1997, Starosta2001}, and
multiple chiral doublets~\cite{MengPRC2006,Ayangeakaa2013,Kuti2014,Liu2016},
and it can handle time-odd fields corresponding to nonzero spatial component
of the vector potential, which is important for understanding these novel
phenomena. Additionally, the CDFT with point-coupling interactions was
developed~\cite{Zhao2010} and applied successfully to magnetic and
antimagnetic rotations~\cite{Meng2016}.

Although the cranking CDFT provides an excellent description of rotational
excitations, the novel effects hidden in the cranking Dirac Hamiltonian have
not been fully revealed. In Refs.~\cite{Guo121,Guo14}, we have applied the
similarity renormalization group (SRG) to decompose the Dirac Hamiltonian
from the relativistic mean field into different components, and obtained the
non-relativistic term, dynamical term, spin-orbital coupling, and Darwin
term. Based on these operators, we have clarified the origin and breaking
mechanism of spin and pseudospin symmetries in spherical nuclei~\cite%
{Guo122,Guo13,Huang2022} as well as deformed nuclei~\cite{Guo15}. Following
our scheme, an analytic expansion to the $1/M^{4}$ order is obtained in Ref.~%
\cite{YXGuo2019} in comparison with the improvement by replacing $M$ with $%
M+S(r)$. A similar work has been done in Ref.~\cite{ZXRen2020}.

In this study, we have combined the cranking CDFT with SRG to disclose the
novel phenomena that may arise in rotating nuclei. By decomposeting the
Hamiltonian from the cranking CDFT, we have obtained three additional
operators: the rotational term, the term relating to Zeeman effect, and the
spin-rotation coupling due to consideration of rotation and spatial
component of vector potential. Based on these operators, we explore the
novel phenomena that may occur in high-speed rotating nuclei, which is
interesting and expects to be observed in experiment. Especially, the spin
rotation coupling is the cause of spin current generation. The spin current,
which is a key concept in the field of spintronics~\cite{Ztic2011}, and can
be employed to probe intriguing properties of quantum materials~\cite%
{Han2019}, and understand the spin Hall effect~\cite{Sinova2015}. It is
worth looking forward to whether this phenomenon occurs in high-speed
rotating nuclei or high-speed rotating neutron stars.

The theoretical formalism is presented in section II. The numerical details
and results are presented in section III. A summary is given in section IV.

\section{Decomposition of cranking Dirac Hamiltonian with similarity
renormalization group}

To investigate the potential emergence of novel phenomena in high-speed
rotating nuclei or neutron stars, we have applied the SRG method to
decompose the Dirac Hamiltonian from the cranking CDFT. Within the cranking
CDFT, the Dirac equation describing rotating nuclei is represented as:
\begin{equation}
H\psi =\varepsilon \psi ,  \label{Diraceq}
\end{equation}%
with the Dirac Hamiltonian
\begin{equation}
H={\vec{\alpha}}\cdot \vec{\pi}+\beta \left( M+S\right) +V-\vec{\Omega}\cdot
\vec{J},  \label{Hamilton}
\end{equation}%
where $\vec{\pi}=\vec{p}-\vec{V}$ and $\vec{J}=\vec{r}\times \vec{\pi}+\vec{%
\Sigma}$ with $\vec{\Sigma}=\frac{\hbar }{4i}{\vec{\alpha}\times \vec{\alpha}%
}$. The scalar potential $S(\vec{r})$ and vector potential $V_{\mu }(\vec{r}%
) $ from the CDFT are represented in the following form:
\begin{eqnarray}
S(\vec{r}) &=&g_{\sigma }\sigma \left( \vec{r}\right) ,  \notag \\
V(\vec{r}) &=&g_{\omega }\omega _{0}\left( \vec{r}\right) +g_{\rho }\tau
_{3}\rho _{30}(\vec{r})+e\frac{1-\tau _{3}}{2}A_{0}(\vec{r}),  \notag \\
\vec{V}(\vec{r}) &=&g_{\omega }\vec{\omega}\left( \vec{r}\right) +g_{\rho
}\tau _{3}\vec{\rho}_{3}(\vec{r})+e\frac{1-\tau _{3}}{2}\vec{A}(\vec{r}).
\end{eqnarray}%
To reveal the underlying physics hidden in the cranking Dirac Hamiltonian,
we transform it into a Schr\"{o}dinger-like form using SRG. The initial
Hamiltonian $H$ is transformed by the unitary operator $U\left( l\right) $
according to
\begin{equation}
H\left( l\right) =U\left( l\right) HU^{\dagger }\left( l\right) ,\text{ \ }%
H(0)=H  \label{unitary}
\end{equation}%
where $l$ is a flow parameter. Differentiation of Eq.(\ref{unitary}) gives
the flow equation as
\begin{equation}
\frac{d}{dl}H\left( l\right) =\left[ \eta \left( l\right) ,H\left( l\right) %
\right] ,  \label{floweq}
\end{equation}%
with the generator $\eta (l)=$ $\frac{dU\left( l\right) }{dl}U^{\dagger
}\left( l\right) $. The choice of $\eta (l)$ is to make $H\left( l\right) $
diagonal in the limit $l\rightarrow \infty $. For the present Hamiltonian,
it is appropriate to choose $\eta (l)$ in the form
\begin{equation}
\eta (l)=\left[ \beta M,H(l)\right] .  \label{generator}
\end{equation}%
In order to solve Eq.(\ref{floweq}), the Hamiltonian $H(l)$ is presented as
a sum of even operator $\varepsilon (l)$ and odd operator $o(l)$:
\begin{equation}
H(l)=\varepsilon (l)+o(l),  \label{hamiltonian}
\end{equation}%
where the even or oddness is defined by the commutation relations of the
respective operators, i.e., $\varepsilon (l)\beta =\beta \varepsilon (l)\ $%
and $o(l)\beta =-\beta o(l)$. The system can be solved perturbatively by the
expansions of $\varepsilon (\lambda )$ and $o(\lambda )$ as follows%
\begin{eqnarray}
\frac{1}{M}\varepsilon \left( \lambda \right) &=&\sum\limits_{i=0}^{\infty }%
\frac{1}{M^{i}}\varepsilon _{i}\left( \lambda \right) ,  \notag \\
\frac{1}{M}o\left( \lambda \right) &=&\sum\limits_{j=1}^{\infty }\frac{1}{%
M^{j}}o_{j}\left( \lambda \right) .  \label{expand}
\end{eqnarray}%
As $\varepsilon (0)=\beta \left( M+S\right) +V-\vec{\Omega}\cdot \vec{J}$
and $o(0)={\vec{\alpha}}\cdot \vec{\pi}$, there are
\begin{eqnarray}
\varepsilon _{0}(0) &=&\beta ,\varepsilon _{1}(0)=\beta S+V-\vec{\Omega}%
\cdot \vec{J},\varepsilon _{i}(0)=0,i\geqslant 2; \\
o_{1}(0) &=&{\vec{\alpha}}\cdot \vec{\pi},o_{i}(0)=0,i\geqslant 2.
\label{initial}
\end{eqnarray}%
With the initial conditions, we have obtained the diagonal Hamiltonian:
\begin{equation}
\varepsilon \left( \infty \right) =M\varepsilon _{0}(\infty )+\varepsilon
_{1}\left( \infty \right) +\frac{1}{M}\varepsilon _{2}(\infty )+\frac{1}{%
M^{2}}\varepsilon _{3}\left( \infty \right) +\cdots
\end{equation}%
The diagonal part describing the positive energy states $H$ can be written
as
\begin{equation}
H=H_{\text{NR}}+H_{\text{Dy}}+H_{\text{SO}}+H_{\text{Dw}}+H_{\text{R}}+H_{%
\text{Z}}+H_{\text{SR}},  \label{hpart}
\end{equation}%
where
\begin{eqnarray}
H_{\text{NR}} &=&\frac{1}{2M}\vec{\pi}^{2}+S+V,  \notag \\
H_{\text{Dy}} &=&-\frac{S}{2M^{2}}\vec{\pi}^{2}+\frac{i\hbar }{2M^{2}}\nabla
S\cdot \vec{\pi},  \notag \\
H_{\text{SO}} &=&+\frac{\hbar }{8M^{2}}\vec{\sigma}\cdot \left[ \vec{\pi}%
\times \left( \vec{E}_{S}+\vec{E}_{V}\right) -\left( \vec{E}_{S}+\vec{E}%
_{V}\right) \times \vec{\pi}\right] ,  \notag \\
H_{\text{Dw}} &=&\frac{\hbar ^{2}}{8M^{2}}\nabla \cdot \left( \vec{E}_{s}-%
\vec{E}_{V}+\vec{E}_{\Omega }\right) ,  \notag \\
H_{\text{R}} &=&-\vec{\Omega}\cdot \vec{j},  \notag \\
H_{\text{Z}} &=&-\frac{\hbar }{2M^{2}}\left( M-S\right) \vec{\sigma}\cdot
\vec{B},  \notag \\
H_{\text{SR}} &=&-\frac{\hbar }{8M^{2}}\vec{\sigma}\cdot \left( \vec{\pi}%
\times \vec{E}_{\Omega }-\vec{E}_{\Omega }\times \vec{\pi}\right) .
\end{eqnarray}%
Here $\vec{E}_{S}=\nabla S$, $\vec{E}_{V}=-\nabla V$, and $\vec{E}_{\Omega
}=\left( \vec{\Omega}\times \vec{r}\right) \times \vec{B}$ with the
assumption $\vec{B}=\nabla \times \vec{V}$. The angular momentum $\vec{j}%
=\left( \vec{r}\times \vec{\pi}+\vec{s}\right) $. $H_{\text{NR}}$
corresponds to the non-relativistic approximation. $H_{\text{Dy}}$ is the
dynamical term. $H_{\text{SO}}$ represents the spin-orbital coupling. $H_{%
\text{Dw}}$ is the Darwin term. $H_{\text{R}}$ is the rotational term. The
Zeeman effect is related to the operator $H_{\text{Z}}$. $H_{\text{SR}}$ is
claimed as the spin-rotation coupling. For the system without rotation, $H_{%
\text{R}}$ and $H_{\text{SR}}$ disappear. From the decomposition of
Hamiltonian, we can explore the important physics hidden in the cranking
Dirac Hamiltonian.

\section{Novel phenomena in rotating nuclei}

\begin{figure}[tbp]
\centering%[!ht]
\includegraphics[width=8.5cm]{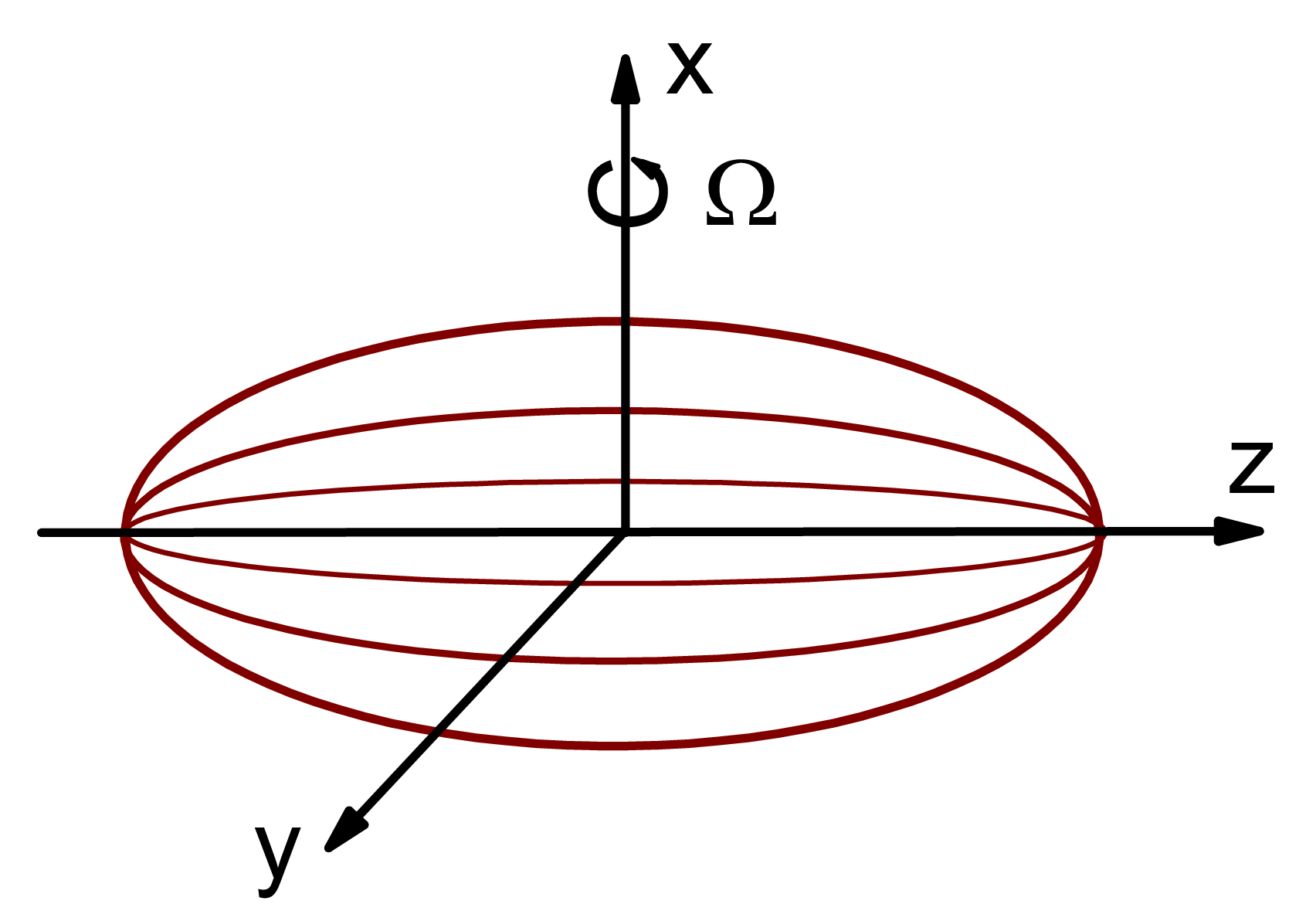}\centering
\caption{(Color online) An axially deformed nucleus rotating around $x$%
-axis. $z$ is its symmetric axis. $\Omega$ is the rotational angular
velocity.}
\label{Fig1}
\end{figure}

Based on the diagonal cranking Dirac Hamiltonian given in Eq.~(\ref{hpart}),
we explore the novel phenomena in rotating nuclei. In Refs.~\cite%
{Guo121,Guo14}, we have probed the physical meanings of these terms $H_{%
\text{NR}}$, $H_{\text{Dy}}$, $H_{\text{SO}}$, and $H_{\text{Dw}}$, and
disclosed their influences on the spin and pseudospin symmetries~\cite%
{Guo122,Guo13,Guo15,Huang2022}. However, since the rotation and the spatial
component of the vector field $\vec{V}(\vec{r})$ are not taken into account
in these references, these terms $H_{\text{R}}$, $H_{\text{Z}}$, and $H_{%
\text{SR}}$ do not appear there, while they have received great attention in
atomic and molecular and condensed matter physics. Hence, it is necessary to
explore the physical effects of these additional terms and their
contributions to the single-particle spectra in atomic nuclei. For
simplicity without losing generality, the nucleus is assumed to be
symmetrical around $z$ axis, the rotation is around $x$-axis, a prolate or
oblate deformation occurs in $z$-axis direction, as shown in Fig.~\ref{Fig1}%
. The nucleus rotates around $x$-axis with the rotational angular velocity $%
\Omega $, i.e., $\vec{\Omega}=\Omega _{x}\vec{e}_{x}$. For the magnetic
field-like $\vec{B}$, only the $x$-direction $\vec{B}=B_{x}\vec{e}_{x}$ is
considered. With the assumption, the Hamiltonian becomes
\begin{eqnarray}
H &=&\frac{\vec{\pi}^{2}}{2M}+\Sigma \left( r\right) -\Omega _{x}j_{x}
\notag \\
&&-\frac{S}{2M^{2}}\vec{\pi}^{2}+\frac{i}{2M^{2}}\nabla S\cdot \vec{\pi}+%
\frac{1}{8M^{2}}\nabla ^{2}\Sigma  \notag \\
&&+\frac{1}{4M^{2}}\vec{\sigma}\cdot \left( \nabla \Delta \times \vec{\pi}%
\right) -\frac{1}{2M^{2}}\left( M-S\right) \sigma _{x}B_{x}  \notag \\
&&+\frac{B_{x}\Omega _{x}}{4M^{2}}+\frac{B_{x}\Omega _{x}}{4M^{2}}\vec{\sigma%
}\cdot \left( \vec{r}_{yz}\times \vec{\pi}\right) .  \label{hsim}
\end{eqnarray}

%\begin{eqnarray}
%H &=&\frac{p^{2}}{2M}+\Sigma \left( r\right) -\Omega J_{x}  \notag \\
%&&-\frac{Sp^{2}}{2M^{2}}+\frac{1}{2M^{2}}\nabla S\cdot \nabla +\frac{1}{%
%8M^{2}}\nabla ^{2}\Sigma   \notag \\
%&&+\frac{1}{4M^{2}}\vec{\sigma}\cdot \left( \nabla \Delta \times \vec{p}%
%\right) -\frac{1}{2M^{2}}\left( M-S\right) \sigma _{x}B  \notag \\
%&&+\frac{B\Omega }{4M^{2}}+\frac{B\Omega }{4M^{2}}\vec{\sigma}\cdot \left(
%\vec{r}_{yz}\times \vec{p}\right)   \label{hsim}
%\end{eqnarray}
To simplify the computation and avoid affecting the conclusion, we have
replaced $\vec{\pi}$ with $\vec{p}$. The energy spectra of the Hamiltonian
in Eq.~(\ref{hsim}) are obtained through basis expansions. It is important
to note that for axisymmetrically deformed nuclei, the third component of
angular momentum is no longer a good quantum number due to rotation. Here,
only axisymmetric quadruple deformation is considered, parity remains a good
quantum number. $^{64}$Ge is chosen as an illustrated example. The
corresponding potentials are adopted as
\begin{eqnarray}
S\left( \vec{r}\right) &=&S_{0}\left( r\right) +S_{2}\left( r\right)
P_{2}\left( \theta \right) ,  \notag \\
V\left( \vec{r}\right) &=&V_{0}\left( r\right) +V_{2}\left( r\right)
P_{2}\left( \theta \right) ,  \label{potential}
\end{eqnarray}%
where $P_{2}\left( \theta \right) =\frac{1}{2}\left( 3\cos ^{2}\theta
-1\right) $. The radial parts in Eq. (\ref{potential}) take a Woods-Saxon
form,
\begin{eqnarray}
S_{0}\left( r\right) &=&S_{\text{WS}}f(r),\text{\ }S_{2}\left( r\right)
=-\beta _{2}S_{\text{WS}}k\left( r\right) ,  \notag \\
V_{0}\left( r\right) &=&V_{\text{WS}}f(r),\text{ }V_{2}\left( r\right)
=-\beta _{2}V_{\text{WS}}k\left( r\right) ,  \label{radialpotential}
\end{eqnarray}%
with $f\left( r\right) =\frac{1}{1+\exp \left( \frac{r-R}{a}\right) }$, and $%
k\left( r\right) =r\frac{df\left( r\right) }{dr}$. Here $V_{\text{WS}}$ and $%
S_{\text{WS}}$ are, respectively, the typical depths of the scalar and
vector potentials in the relativistic mean field chosen as 350 and -405 MeV,
the diffuseness of the potential $a$ is fixed as 0.67 fm, and $\beta _{2}$
is the axial deformation parameter of the potential. The radius $R\equiv
r_{0}A^{1/3}$ with $r_{0}=1.27$ fm. With these parameters are determined,
the energy spectra of Hamiltonian $H$ can be calculated.
\begin{figure}[tbp]
\centering
\includegraphics[width=8.cm]{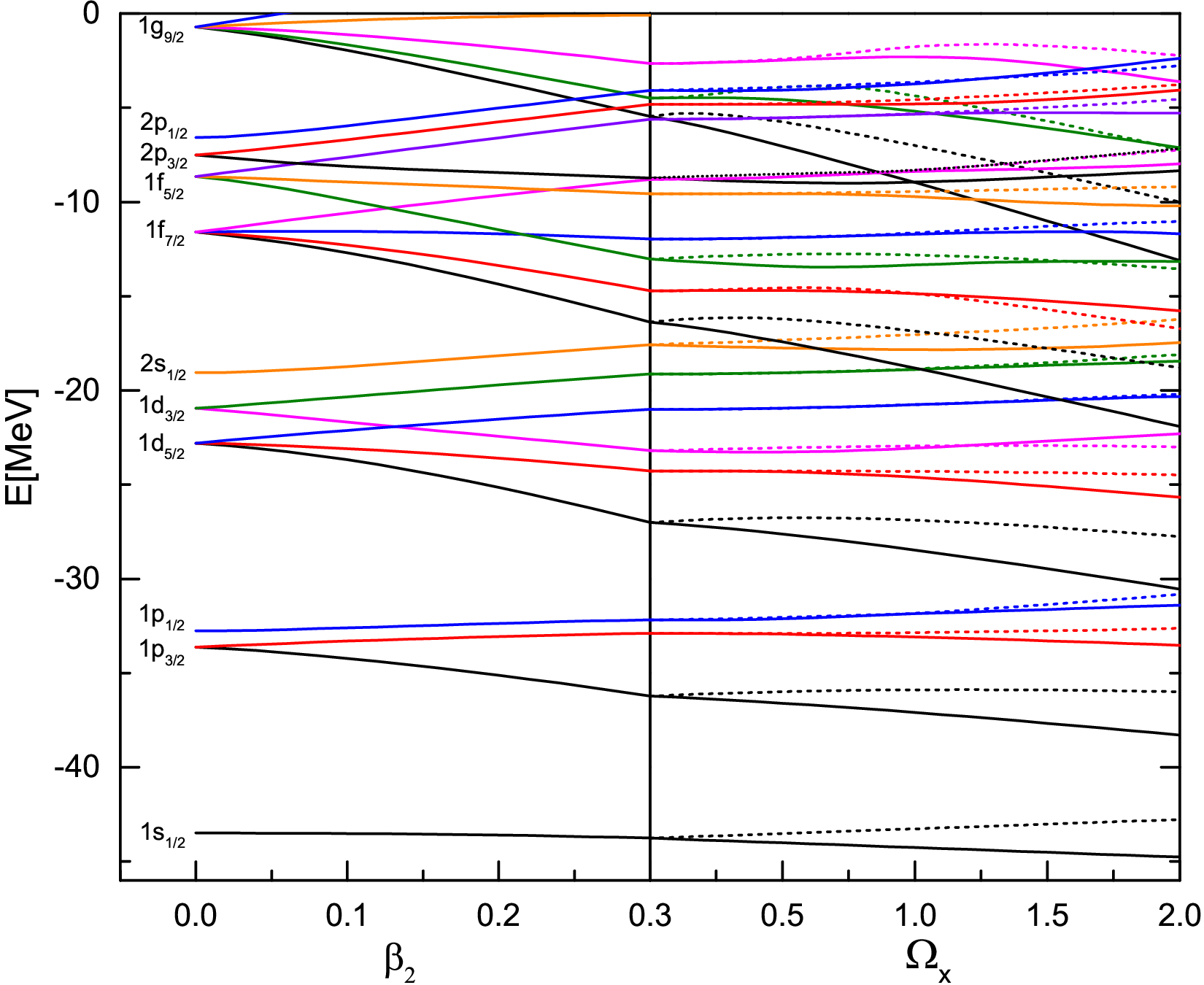}\centering
\caption{(Color online) The single-particle spectra and their evolutions to
the quadruple deformation $\protect\beta _{2}$ and the rotational angular
velocity around $x$-axis $\Omega _{x}$.}
\label{Fig2}
\end{figure}

The single-particle spectra and their evolutions with the quadruple
deformation $\beta _{2}$ and the rotational angular velocity around the $x$%
-axis $\Omega _{x}$ are presented in Fig.~\ref{Fig2}. The left panel shows
the Nilsson levels with deformation $\beta _{2}$ ranging from 0 to 0.3. The
right panel illustrates the variation of single-particle spectra with $%
\Omega _{x}$ for $\beta _{2}=0.3$. From the spherical to axial deformation,
the levels experience splitting, as the well-known Nilsson levels are
clearly visible. For axially deformed nuclei, there are signature splittings
in doubly degenerate Nilsson levels. Compared to unaligned spin states, the
signature splitting is larger for spin aligned states. In particular, for
the spin aligned state with the least third component of total angular
momentum, the signature splitting is the largest and most sensitive to $%
\Omega _{x}$ as shown in Fig.~\ref{Fig2} for the single particle level $%
1g_{9/2,1/2}$. The single-particle energy drops faster with increasing $%
\Omega _{x}$ due to the stronger influence of the Coriolis force on the
energy spectrum, which is consistent with the prediction of the cranking
shell model. Signature splitting is an important phenomenon in rotating
nuclei. Our present calculations based on the diagonal Dirac Hamiltonian in
Eq.~(\ref{hsim}) can reproduce the signature splitting, which represents the
first time that signature splitting in nuclei has been understood in a
relativistic framework.
\begin{figure}[tbp]
\centering
\includegraphics[width=8.5cm]{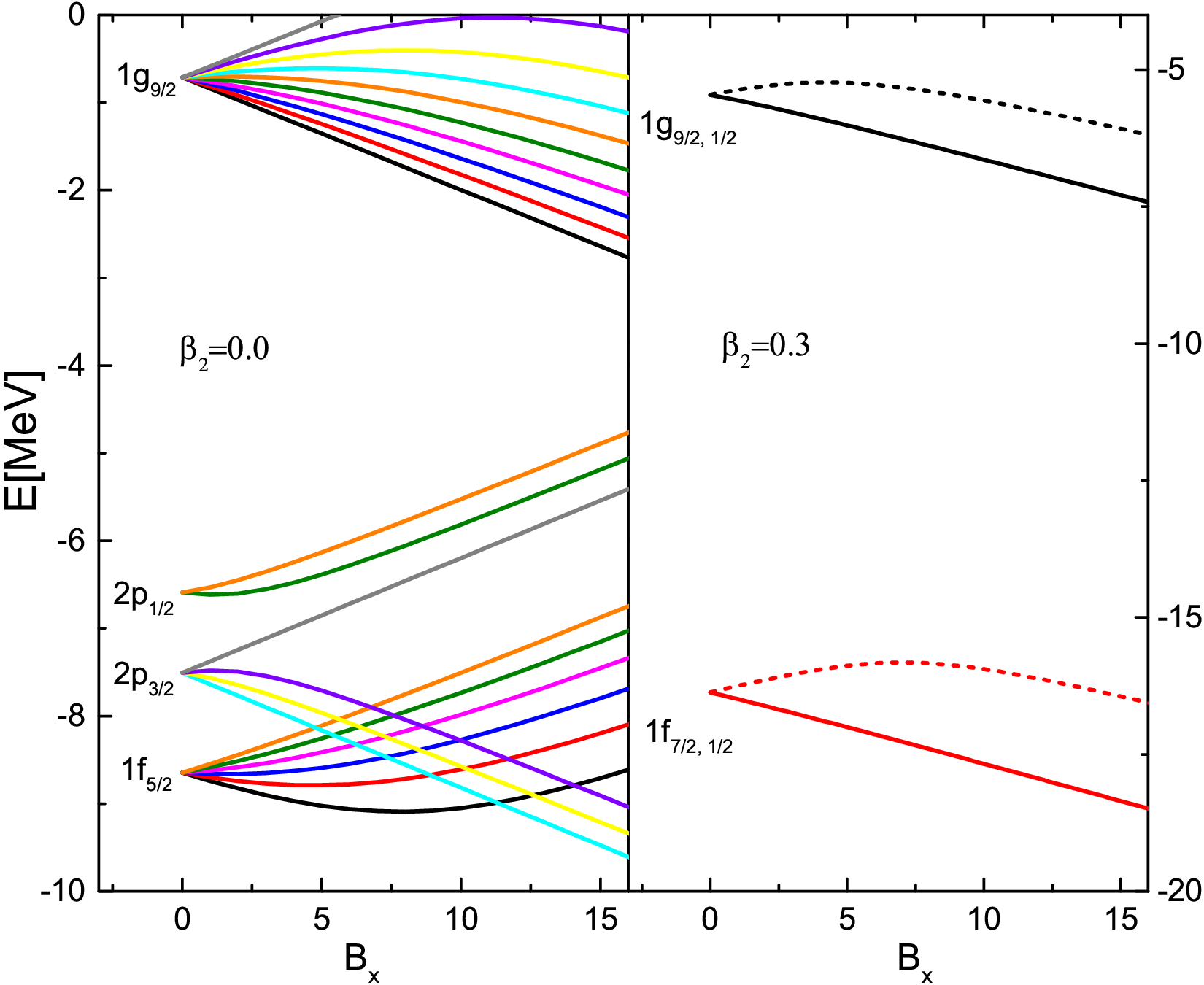}\centering
\caption{(Color online) The single-particle spectra and their evolutions to
the magnetic field-like $B_{x}$ for the spherical nuclei and deformed nuclei
with $\protect\beta _{2}=0.3$.}
\label{Fig3}
\end{figure}

In addition to the signature splitting caused by this cranking item $H_{%
\text{R}}$, the additional $H_{\text{Z}}$ deserves special attention. It is
the cause of the Zeeman effect in the atomic and molecular physics. To
understand the role of $H_{\text{Z}}$ in atomic nuclei, the single-particle
spectra and their evolutions to the magnetic field-like $B_{x}$ for the
spherical nuclei and deformed nuclei with $\beta _{2}=0.3$ are presented in
Fig.~\ref{Fig3}. In the left panel, the result for spherical nuclei is
shown, while in the right panel, the results for deformed nuclei with $\beta
_{2}=0.3$ are shown. When $B_{x}\neq 0$, every degenerate spherical level
with the total angular momentum $j$ is split into $2j+1$ levels. As $B_{x}$
increases, the level split increases as well. For deformed nuclei, the
doubly degenerate Nilsson levels split when $B_{x}\neq 0$. The some split
are remarkable, while others are considerably small or even unobservable.
However, for the spin aligned state with the least third component of total
angular momentum, the level split is significant and sensitive to $B_{x}$,
as seen in the right panel in Fig.~\ref{Fig3} for $1f_{7/2,1/2}$ and $%
1g_{9/2,1/2}$. For the convenience of describing the problem, the splitting
caused by the $H_{\text{Z}}$ is claimed as Zeeman splitting-like. When $B_{x}$ is
large enough, the Zeeman splitting-like becomes considerable and may lead to an
observable Zeeman effect-like. The observation of Zeeman effect-like can help us to
understand the level structure and the strength of the magnetic field-like $%
B_{x}$. It is well worth experimentally probing the Zeeman effect-like in the
nuclei.
\begin{figure}[tbp]
\centering%[!ht]
\includegraphics[width=8.5cm]{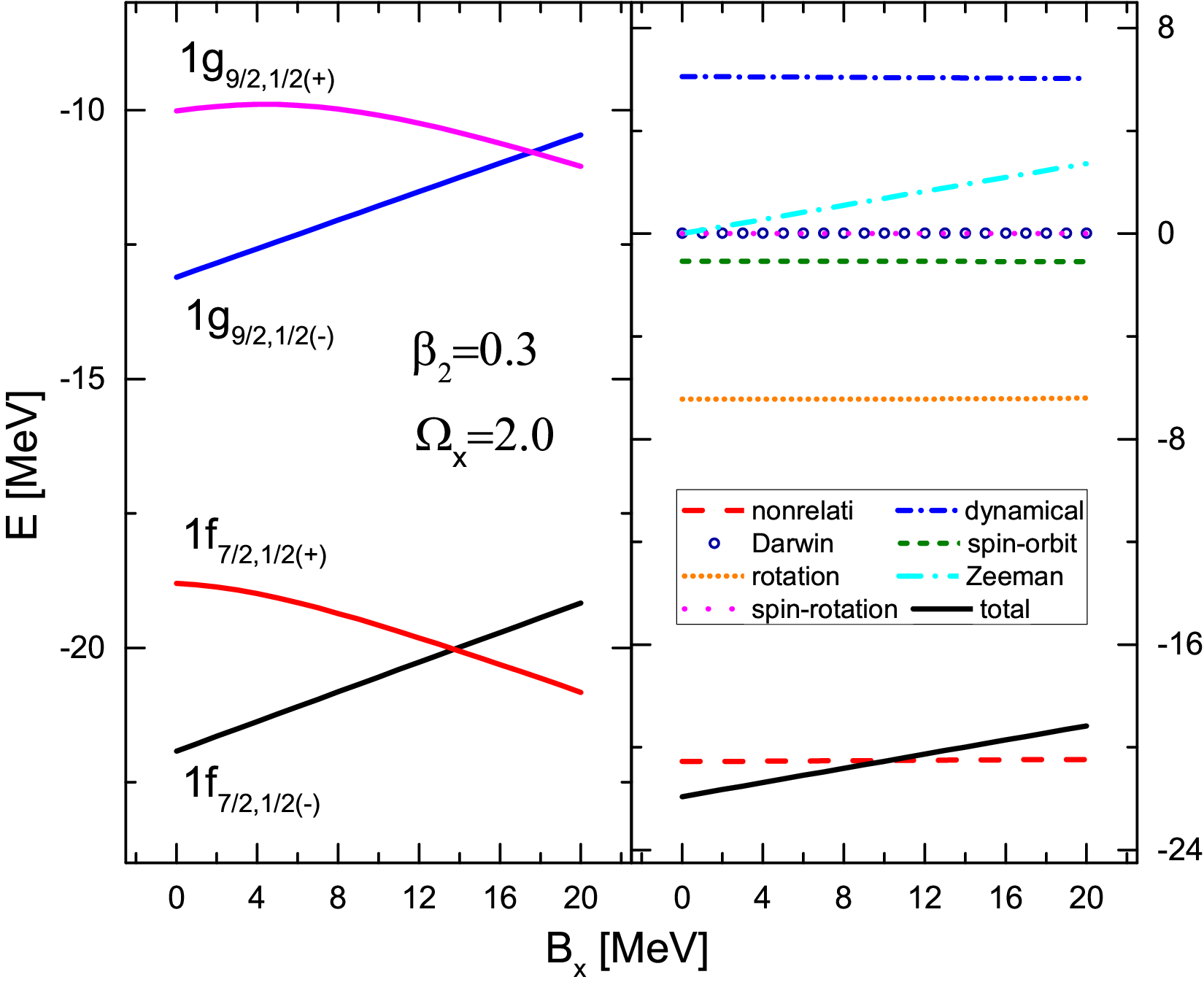}\centering
\caption{(Color online) The energies of single particle levels and their
components evolve to the magnetic field-like $B_{x}$ for the rotational
nuclei with $\protect\beta _{2}=0.3$ and $\Omega _{x}=2.0$ MeV. The left
panel shows the signature doublets $1f_{7/2,1/2}$ and $1g_{9/2,1/2}$. The
right panel illustrates the contributions of every component to the
single-particle level $1f_{7/2,1/2(-)}$.}
\label{Fig4}
\end{figure}

Moreover, we have investigated the influence of the magnetic field-like $%
B_{x}$ on signature splitting in rotating nuclei with $\beta _{2}=0.3$ and $%
\Omega _{x}=2.0$ MeV. The left panel of Fig. \ref{Fig4} displays the level
splittings for the signature doublets $1f_{7/2,1/2}$ and $1g_{9/2,1/2}$ and
their evolutions to $B_{x}$. As $B_{x}$ increases, the levels for the $%
1f_{7/2,1/2(-)}$ and $1g_{9/2,1/2(-)}$ go up, while those for the $%
1f_{7/2,1/2(+)}$ and $1g_{9/2,1/2(+)}$ go down. There are also the levels of
the signature doublet crosses at a specific $B_{x}$. To clarify the
evolution of level splittings to $B_{x}$, the contributions of every
component to the single-particle level $1f_{7/2,1/2(-)}$ are displayed in
the right panel of Fig.~\ref{Fig4}. The non-relativistic, dynamical, and
rotational terms have significant contributions to the total energy. The
spin-orbit coupling and the term relating to Zeeman effect-like contribute
relatively little to the total energy. The Darwin term and the spin-rotation
coupling contribute very little to the total energy. Except for the term
relating to Zeeman effect-like, the other contributions to the total energy are
independent of $B_{x}$. The energy increasing with $B_{x}$ comes entirely
from the contribution of the term relating to Zeeman effect-like, which is the
cause of Zeeman effect-like in rotating nuclei.
\begin{figure}[tbp]
\centering%[!ht]
\includegraphics[width=8.5cm]{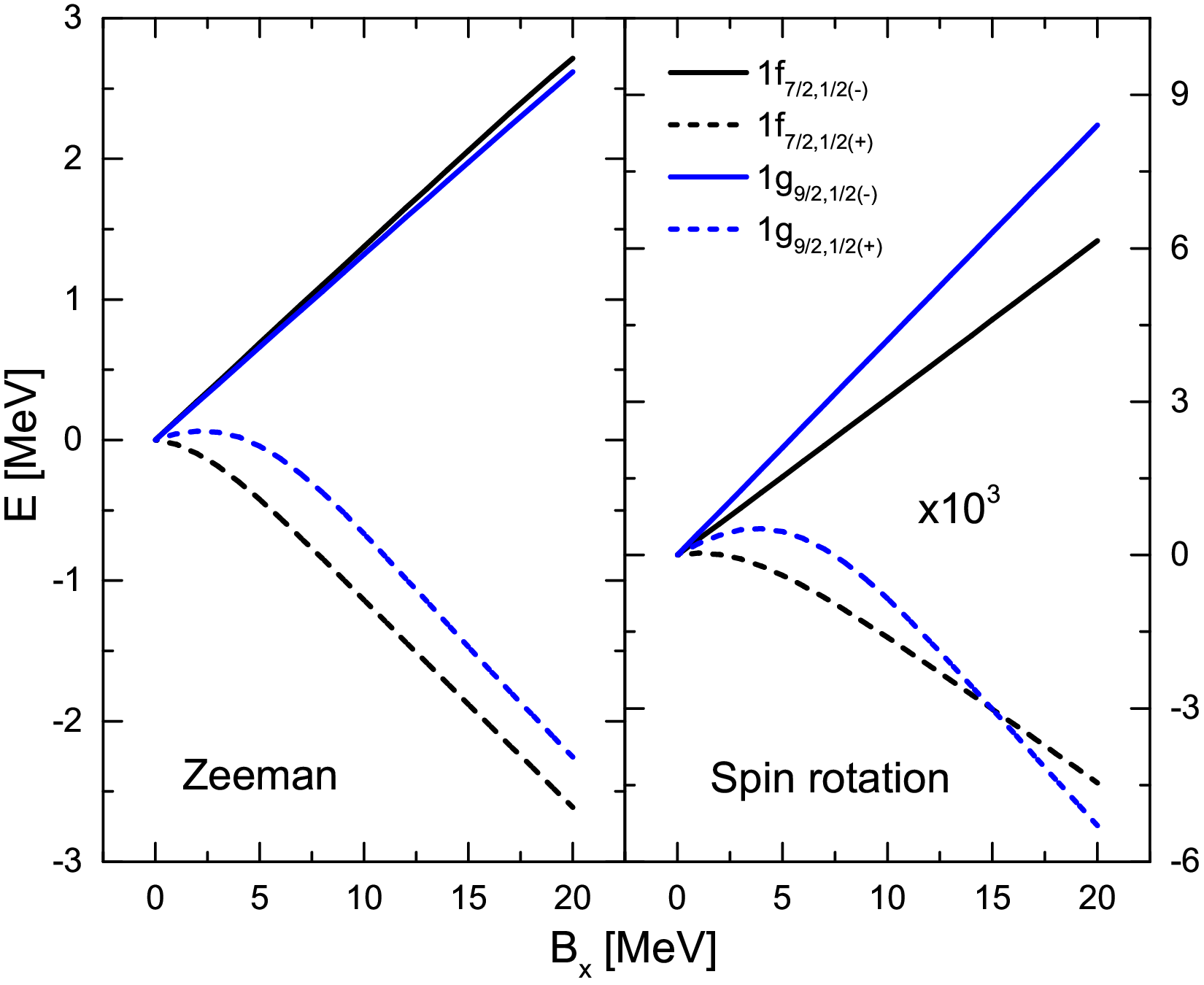}\centering
\caption{(Color online) The energies contributed by the operator relating to
Zeeman effect-like and the spin-rotation coupling, and their evolutions with the
magnetic field-like $B_{x}$ in the single particle levels $1f_{7/2,1/2(-)}$,
$1f_{7/2,1/2(+)}$, $1g_{9/2,1/2(-)}$, and $1g_{9/2,1/2(+)}$. The other
physical quantities are the same as Fig.~\protect\ref{Fig4}.}
\label{Fig5}
\end{figure}

To better understand the Zeeman effect-like in rotating nuclei, the left panel of
Fig.~\ref{Fig5} displays the signature splittings contributed by the term
relating to Zeeman effect-like for the two signature doublets $1f_{7/2,1/2}$ and $%
1g_{9/2,1/2}$ with $\beta _{2}=0.3$ and $\Omega _{x}=2.0$ MeV. With the
increase of $B_{x}$, the energy increases for the $1f_{7/2,1/2(-)}$ and $%
1g_{9/2,1/2(-)}$, while the energy decreases for the $1f_{7/2,1/2(+)}$ and $%
1g_{9/2,1/2(+)}$. When $B_{x}=20$ MeV, the increasing or reducing energy is
about 3.0 MeV, which is quite observable. These indicate that the Zeeman
effect-like plays a significant role in the signature splitting in rotating
nuclei. For comparison, the spin-rotation coupling effect is displayed in
the right panel in Fig.~\ref{Fig5}. The trend with $B_{x}$ is similar, but
the magnitudes are much smaller than 3 orders of magnitude.

\section{Spin current}

Although the spin-rotation coupling contributes less to the signature
splitting, it may cause the occurrence of spin current. To investigate the
spin current, we extract the operator of spin-rotation coupling from the
total Hamiltonian as follows:
\begin{equation}
H_{\text{SR}}=-\frac{\hbar }{8M^{2}}\vec{\sigma}\cdot \left( \vec{\pi}\times
\vec{E}_{\Omega }-\vec{E}_{\Omega }\times \vec{\pi}\right) .
\end{equation}%
By the definition, the speed relating to the spin-rotation coupling is
\begin{eqnarray}
\vec{v}_{s} &=&\frac{1}{i\hbar }\left[ \vec{r},H_{\text{SR}}\right]  \notag
\\
&=&\frac{1}{i\hbar }\left[ \vec{r},\frac{\hbar }{8M^{2}}\vec{\sigma}\cdot
\left( i\hbar \nabla \times \vec{E}_{\Omega }+2\vec{E}_{\Omega }\times \vec{%
\pi}\right) \right]  \notag \\
&=&\frac{1}{4iM^{2}}\left[ \vec{r},\vec{\sigma}\cdot \left( \vec{E}_{\Omega
}\times \vec{p}\right) \right] .
\end{eqnarray}%
If the only $\vec{\Omega}=\Omega _{x}\vec{e}_{x}$ is considered, there is
\begin{equation}
\vec{E}_{\Omega }=\Omega _{x}\left[ -\left( B_{y}y+B_{z}z\right) \vec{e}%
_{x}+B_{x}y\vec{e}_{y}+B_{x}z\vec{e}_{z}\right] .
\end{equation}%
The speed relating to the spin-rotation coupling is
\begin{eqnarray}
\vec{v}_{s} &=&\frac{\hbar \Omega _{x}}{4M^{2}}\Big[\left( \sigma
_{y}z-\sigma _{z}y\right) B_{x}\vec{e}_{x}  \notag \\
&&-\left( \sigma _{z}B_{y}y+\sigma _{z}B_{z}z+\sigma _{x}B_{x}z\right) \vec{e%
}_{y}  \notag \\
&&+\left( \sigma _{x}B_{x}y+\sigma _{y}B_{y}y+\sigma _{y}B_{z}z\right) \vec{e%
}_{z}\Big]
\end{eqnarray}%
Due to rotation in the $x$-direction, only $\vec{B}=B_{x}\vec{e}_{x}$ is
dominant.
\begin{equation}
\vec{v}_{s}=\frac{\hbar \Omega _{x}B_{x}}{4M^{2}}\left[ \left( \sigma
_{y}z-\sigma _{z}y\right) \vec{e}_{x}+\sigma _{x}\left( y\vec{e}_{z}-z\vec{e}%
_{y}\right) \right]
\end{equation}%
\begin{eqnarray}
v_{x//}^{s} &=&\frac{B_{x}\Omega _{x}}{4M^{2}}\left( \sigma _{y}z-\sigma
_{z}y\right) \vec{e}_{x}, \\
v_{x\bot }^{s} &=&\frac{B_{x}\Omega _{x}}{4M^{2}}\sigma _{x}\left( y\vec{e}%
_{z}-z\vec{e}_{y}\right) =\frac{B_{x}\Omega _{x}}{4M^{2}}\sigma _{x}r_{yz}%
\vec{e}_{\phi }^{yz}.
\end{eqnarray}%
The spin current is defined as
\begin{eqnarray}
J_{s} &=&M_{r_{yz}}Tr\sigma _{x}v_{x\bot }^{s}=M_{r_{yz}}\frac{B_{x}\Omega
_{x}}{2M^{2}}r_{yz}\vec{e}_{\phi }^{yz}, \\
\left\vert J_{s}\right\vert &=&M_{r_{yz}}\frac{B_{x}\Omega _{x}}{2M^{2}}%
r_{yz}.
\end{eqnarray}%
For a single nucleon populating at $r_{yz}$, an estimate of the size of the
spin current
\begin{equation}
\left\vert J_{s}\right\vert =\frac{B_{x}\Omega _{x}}{2M}r_{yz}.
\end{equation}%
Assuming $\Omega _{x}=2$ MeV, $B_{x}=20$ MeV, and $r_{yz}=5$ fm, the
calculated spin current $\left\vert J_{s}\right\vert \approx 5.41\times
10^{-4}$ MeV $=0.541$ keV. With the advancement of laser control technology,
we can prepare strong enough lasers to control nuclear reactions and nuclear
structures. It may be possible to regulate nucleon spin and spin current in
the near future.

\section{Summary}

In summary, we have combined the cranking covariant density functional
theory with similar renormalization group to decompose the Dirac Hamiltonian
from the cranking CDFT into several Hermit components, including the
non-relativistic term, the dynamical term, the spin-orbit coupling, and the
Darwin term. Especially, we have obtained the rotational term, the term
relating to Zeeman effect-like, and the spin-rotation coupling due to the
consideration of rotation and spatial component of vector potential. Based
on these operators, we have explored the novel phenomena that may occur in
rotating nuclei.

One of the most fascinating phenomena is signature splitting. It is shown
that the signature splitting is larger for spin aligned states than that for
unaligned spin states. Especially, the spin aligned state with the least
third component and the largest total angular momentum experiences the
largest signature splitting. The energy of this state decreases faster with
the increase of $\Omega _{x}$, which is attributed to the increased strength
of the Coriolis force.

In addition to the signature splitting, we also investigated the level
splitting caused by the operator relating to Zeeman effect-like. For spherical
nuclei, the degenerate level is split into $2j+1$ levels, and the level
splitting increases with the increase of magnetic field-like $B_{x}$. For
deformed nuclei, the doubly degenerate levels split. For the spin aligned
state with the least third component, the level splitting is remarkable and
sensitive to $B_{x}$. When $B_{x}$ is sufficiently large, the Zeeman
splitting-like becomes significant enough to produce an observable Zeeman effect-like
for both spherical and deformed nuclei. The observation of the Zeeman effect-like
can provide valuable insights in the level structure and the strength of the
magnetic field-like $B_{x}$ in rotating nuclei.

Moreover, we have investigated the influence of the magnetic field-like $%
B_{x}$ on signature splitting in rotating nuclei with $\beta _{2}=0.3$ and $%
\Omega _{x}=2.0$ MeV. We observed that the levels with the signature (-) go
up while those with the signature (+) go down with the increase of $B_{x}$.
The trend of the level energies with $B_{x}$ is fully attributed to the
contribution of the term related to Zeeman effect. Although these terms,
such as the non-relativistic, the dynamical, the rotational, and the
spin-orbit coupling, contribute considerably to the signature splittings,
they are almost independent of $B_{x}$. Furthermore, we have extracted the
signature splitting contributed by the term relating to Zeeman effect. With
the increase of $B_{x}$, the energy increase for the state with signature
(-) and the energy decrease for the state with signature (+) are
considerably observable. These imply the operator relating to Zeeman effect
play an important role at signature splitting in rotating nuclei.

Based on the spin-rotation coupling, we calculate the spin current, evaluate
the magnitude of spin current, and point out the possibility of regulating
the nuclear spin and spin current in the near future with advances in laser
control technology.

The theoretical predictions on these novel phenomena have significant
reference value for experimental detection of such phenomena in atomic
nuclei.

\section{Acknowledgments}

This work was partly supported by the National Natural Science Foundation of
China under Grants No. 11935001 and No.11575002, Anhui project (Z010118169),
Heavy Ion Research Facility in Lanzhou (HIRFL) (HIR2021PY007), the project
of Key Laboratory of High Precision Nuclear Spectroscopy in Chinese Academy
of Sciences, and the Graduate Research Foundation of Education Ministry of
Anhui Province under Grant No. YJS20210096.

%\end{CJK*}

\end{document}